\documentclass[
    ,final            
  ]
  {aipproc}

\layoutstyle{8x11single}

\def\gsim{\mathrel{\raise.3ex\hbox{$>$\kern-.75em\lower1ex\hbox{$\sim$}}}}
\def\lsim{\mathrel{\raise.3ex\hbox{$<$\kern-.75em\lower1ex\hbox{$\sim$}}}}

\begin{document}

\title{Early Universe cosmology with mirror dark matter}

\classification{26.35.+c, 95.30.Cq, 95.35.+d, 98.65.Dx, 98.70.Vc, 98.80.-k}
\keywords      {dark matter, structure formation, Big Bang nucleosynthesis, cosmic microwave background, large scale structure}

\author{P.~Ciarcelluti}{
  address={IFPA, D\'epartement AGO, Universit\'e de  Li\`ege, B-4000 Li\`ege, Belgium}
}

\begin{abstract}
Mirror matter is a stable self-collisional dark matter candidate. 
If exact mirror parity is a conserved symmetry of nature, there could exist a parallel hidden (mirror) sector of the Universe which has the same kind of particles and the same physical laws of our (visible) sector. 
The two sectors interact each other predominantly via gravity, therefore mirror matter is naturally ``dark''. 
Here I briefly review the cosmological signatures of mirror dark matter, as Big Bang nucleosynthesis, primordial structure formation and evolution, cosmic microwave background and large scale structure power spectra, together with its compatibility with the interpretation of the DAMA annual modulation signal in terms of photon--mirror-photon kinetic mixing.
Summarizing the present status of research and comparing theoretical results with observations/experiments, it emerges that mirror matter is not just a viable, but a promising dark matter candidate.
\end{abstract}

\maketitle


\section{Introduction}

Mirror matter is a stable self-interacting dark matter candidate that emerges if one, instead of (or in addition to) assuming a symmetry between bosons and fermions (supersymmetry), assumes that nature is parity symmetric.
It is a matter of fact that the weak nuclear force is not parity symmetric.
The main theoretical motivation for the mirror matter hypothesis is that it constitutes the simplest way to restore parity symmetry in the physical laws of nature. 
When Lee and Yang proposed the non-parity of weak interactions in 1956, they mentioned also the possibility to restore parity by doubling the number of particles in the Standard Model \cite{Lee:1956qn}.
Thereby the Universe is divided into two sectors with the same particles and interactions, but opposite handedness, that interact mainly by gravity. 
In the minimal parity-symmetric extension of the Standard Model \cite{Foot:1991bp,Pavsic:1974rq}, the group structure is $G\otimes G'$, where $G=SU(3) \otimes SU(2) \otimes U(1)$ and the prime $(')$ denotes, as usual, the mirror sector.
In this model the two sectors are described by the same lagrangians, but where ordinary particles have left-handed interactions, mirror particles have right-handed interactions. 
Besides gravity, mirror matter could interact with ordinary matter via the so-called kinetic mixing of gauge bosons, or via unknown fields that carry both ordinary and mirror charges. 
If such interactions exist they must be weak and are, therefore, negligible for many cosmological processes, but could be determinant for the detection of mirror dark matter in some contexts.
Since photons do not interact with mirror baryons, or interact only via the weak kinetic mixing, mirror matter constitutes a natural candidate for the dark matter in the Universe.

Many consequences of mirror matter for particle physics and astrophysics have been studied during the last decades.
The reader can refer to \cite{Okun:2006eb} for a review of the history of mirror matter and a list of most relevant papers published before 2006.

Like their ordinary counterparts, mirror baryons can form atoms, molecules and astrophysical objects such as planets, stars and globular clusters.
However, even though the microphysics is the same in both sectors, the initial conditions, and then the macrophysics, of the mirror sector should be different, because the cosmology must be different. 
In particular, Big Bang nucleosynthesis (BBN) requires that the mirror sector has a lower temperature than the ordinary one \cite{Berezhiani:1995am,Berezhiani:2000gw}. 
This has implications for the thermodynamics of the early Universe \cite{Berezhiani:2000gw,Ciarcelluti:2008vs} and for the key cosmological epochs.
Analytical results and numerical calculations of BBN \cite{Berezhiani:2000gw,Ciarcelluti:2008vm}
show that the mirror sector should be helium dominated, and the abundance of heavy elements is expected to be higher than in the ordinary sector.
Invisible stars made of mirror baryons are candidates for Massive Astrophysical Compact Halo Objects (MACHOs), which have been observed via microlensing events \cite{Blinnikov:1996fm,Foot:1999hm,Mohapatra:1999ih}.
Mirror stars contain more helium and less hydrogen, and therefore have different properties than ordinary stars \cite{Berezhiani:2005vv}.
The accretion of mirror matter onto celestial objects 
\cite{Blinnikov:1983gh,Khlopov:1989fj,Blinnikov:1982eh} and the presence of mirror matter inside compact stars, as for example neutron stars, could have interesting observable effects \cite{Sandin:2008db,Blinnikov:2009nn}.
The consequences of mirror matter on primordial structure formation, cosmic microwave background (CMB) and large scale structure (LSS) distribution of matter have been studied 
\cite{Ignatiev:2003js,Berezhiani:2003wj,Ciarcelluti:2003wm,Ciarcelluti:2004ij,
Ciarcelluti:2004ik,Ciarcelluti:2004ip}.
These studies provide stringent bounds on the mirror sector and prove that it is a viable candidate for dark matter. 
In addition, mirror matter provides one of the few potential explanations for the recent DAMA/LIBRA annual modulation signal \cite{Bernabei:2008yi,Foot:2008nw,Ciarcelluti:2008qk,Foot:2009mw}, and possibly other experiments, as suggested for low energy electron recoil data from the CDMS collaboration \cite{Foot:2009gk}.

If the mirror (M) sector exists, then the Universe along with the ordinary (O) particles should contain their mirror partners, but their densities are not the same in both sectors.
In fact, the BBN bound on the effective number of extra light neutrinos implies that the M sector has a temperature lower than the O one, that can be naturally achieved in certain inflationary models \cite{Berezhiani:1995am}.
Then, two sectors have different initial conditions, they do not come into thermal equilibrium at later epoch and they evolve independently, separately conserving their entropies, and maintaining approximately constant the ratio among their temperatures.

All the differences with respect to the ordinary world can be described in terms of only two free parameters: 
\begin{eqnarray}\label{mir-param}
x \equiv \left( s' \over s \right)^{1/3} \approx {T' \over T}
~~~~~~ {\rm and} ~~~~~~
\beta \equiv {\Omega'_{\rm b} \over \Omega_{\rm b}} ~~,
\end{eqnarray}
where $T$ ($T'$), $\Omega_{b}$ ($\Omega'_{b}$), and $s$ ($s'$) are respectively the ordinary (mirror) photon temperature, cosmological baryon density, and entropy density.
The lightest bounds on the mirror parameters are $ x < 0.7 $ and $ \beta > 1 $, the first one coming from the BBN limit and the second one from the hypothesis that a relevant fraction of dark matter is made of mirror baryons.

As far as the mirror world is cooler than the ordinary one, $x < 1$, in the mirror world all key epochs (as are baryogenesis, nucleosynthesis, recombination, etc.) proceed in somewhat different conditions than in ordinary world.
Namely, in the mirror world the relevant processes go out of equilibrium earlier than in the ordinary world, which has many far going implications. 

In this paper I consider the cosmological effects of the gravitational interaction between O and M matter, and I add the effects of photon--mirror-photon kinetic mixing only when showing the compatibility of the mirror dark matter interpretation of DAMA/LIBRA annual modulation signal with the present cosmological scenario.

\section{Thermodynamics of the early Universe}

In general, during most history of the Universe, we can approximate
\begin{equation} \label{xmir}
  x \equiv \left( s' \over s \right)^{1/3}
    = \left[q'(T') \over q(T) \right]^{1/3}{T' \over T} \approx {T' \over T} ~~,
\end{equation}
where now $q(T)$ and $q'(T)$ are respectively the O and M entropic degrees of freedom, and the M photon temperature $T'(T)$ is a function of the O one.
This approximation is valid only if the temperatures of the two sectors are not too different, or otherwise if we are far enough from crucial epochs, like $e^+$-$e^-$ annihilation.
When we investigate the range of temperatures interested by this phenomenon, as for example studying the BBN case, we need to exactly compute the trends of the thermodynamical quantities \cite{Ciarcelluti:2008vs}.

Since ordinary and mirror sectors have the same microphysics, we may consider that the neutrino decoupling temperature $T_{D\nu}$ is the same in both of them, that is $T_{D\nu}=T_{D\nu}'$. 
We use this fact, together with the entropy conservation, to find equations which will give the mirror photon temperature $T'$ and the ordinary and mirror thermodynamical quantities corresponding to any values of the ordinary photon temperature $T$. From them it is possible to work out the total effective number of degrees of freedom (DOF) in both sectors, which can be, as common in the literature, expressed in terms of total effective number of neutrinos $ N_\nu $.
The presence of the other sector, indeed, leads in both sectors to the same effects of having more particles. 

At temperatures we are interested in (below $\sim$ 10 MeV) we can in general use the following equations derived from the conservations of separated entropies:
\begin{eqnarray}\label{eqs:1-2}
\frac{22}{21}&=&\frac{\frac{7}{8}q_{e}(T')+q_{\gamma}}{\frac{7}{8}q_{\nu}} \left(\frac{T'}{T_{\nu}'}\right)^3 ~~; \nonumber\\
\frac{22}{21}&=&\frac{\frac{7}{8}q_{e}(T)+q_{\gamma}}{\frac{7}{8}q_{\nu}} \left(\frac{T}{T_{\nu}}\right)^3 ~~,
\end{eqnarray}
and neglecting the entropy exchanges between the sectors (imposing $x$ constant):
\begin{equation}\label{eq:3}
x^3 = \frac{s'\cdot a^3}{s \cdot a^3} =
\frac{\left[ \frac{7}{8}q_e(T')+ q_{\gamma} \right] T'^3+ \frac{7}{8} q_{\nu} 
T_{\nu}'\,^3} {\left[  \frac{7}{8}q_e(T)+ q_{\gamma} \right] T^3 + 
\frac{7}{8} q_{\nu} T_{\nu}^3} ~~,
\end{equation}
where $s$ is the entropy density, $a$ the scale factor, $q_i$ the entropic DOF of species $i$.

At $T\simeq T_{D\nu}$ ($T'_{D\nu}$) ordinary (mirror) neutrinos decouple and soon after ordinary (mirror) electrons and positrons annihilate.

The mirror world must be colder than the ordinary one and therefore the neutrino decoupling takes place before in the mirror sector.
We can split the early Universe evolution in three phases.

\noindent 1) $T > T_{D\nu'}$:
photons and neutrinos are in thermal equilibrium in both worlds, that is $T_{\nu} = T \; , \; T_{\nu}' = T'$. 

\noindent 2) $T_{D\nu} < T \leq T_{D\nu'}$:
at $T \simeq T_{D\nu'}$ M neutrinos decouple and soon after M electrons and positrons annihilate, raising the M photon temperature ($T' \neq T_{\nu}'$). 
Nevertheless, O photons and neutrinos still have the same temperature ($T = T_{\nu}$).

\noindent 3) $T \leq T_{D\nu}$:
at $T\simeq T_{D\nu}$ O neutrinos decouple and soon after O electrons and positrons annihilate, raising the O photon temperature ($T \neq T_{\nu}$).

Equations \eqref{eqs:1-2} and \eqref{eq:3} are solved numerically in order to work out the total, ordinary and mirror numbers of entropic ($q$) and energetic ($g$) DOF at any temperature $T$. 
The corresponding number of neutrinos $N_{\nu}$ is found assuming that all particles contributing to the Universe energy density, to the exclusion of electrons, positrons and photons, are neutrinos; in formula this means
\begin{equation}\label{Nnu}
N_{\nu} = \frac{\bar g - g_{e^{\pm}}(T) - g_{\gamma}}{\frac{7}{8}\cdot 2} \cdot \left( \frac{T}{T_{\nu}} \right)^4 ~~.
\end{equation}

In Table \ref{Nnu-ord} I report the asymptotic numerical results before ($ T $ = 5 MeV) and after ($ T $ = 0.005 MeV) BBN for different values of $x$. 
I stress that the standard value $N_{\nu} = 3$ is the same at any temperatures, while a distinctive feature of the mirror scenario is that the number of neutrinos raises with temperature and with $x$.
Anyway, this effect is not a problem; on the contrary it may be useful since recent data fits give indications for a number of neutrinos at recent times higher than at BBN.
\begin{table}
\begin{tabular}{lrrrrrr}
\hline
   \tablehead{1}{r}{b}{T(MeV)}
  & \tablehead{1}{r}{b}{standard}
  & \tablehead{1}{r}{b}{$\mathbf{x=0.1}$}
  & \tablehead{1}{r}{b}{$\mathbf{x=0.3}$}
  & \tablehead{1}{r}{b}{$\mathbf{x=0.5}$}
  & \tablehead{1}{r}{b}{$\mathbf{x=0.7}$}   \\
\hline
5 & 3.00000 & 3.00063 & 3.04989 & 3.38430 & 4.47563\\
0.005 & 3.00000 & 3.00074 & 3.05997 & 3.46270 & 4.77751\\
\hline
\end{tabular}
\caption{Effective $ N_\nu $ in the ordinary sector.}
\label{Nnu-ord}
\end{table}

It is possible to approximate the difference between the effective numbers of neutrinos before and after BBN process with the following expression:
\begin{equation}\label{N_nu_mir_approx}
N_{\nu} (T \ll T_{ann \, e^{\pm}}) - N_{\nu} (T \gg T_{D\nu})
= x^4 \cdot \frac{1}{\frac{7}{8}\cdot 2} \left[ 10.75 - 3.36 \left( \frac{11}{4} \right)^\frac{4}{3}\right]
\simeq 1.25 \cdot x^4 ~~.
\end{equation}

The effective number of neutrinos in the mirror sector can be worked out in a similar way, and the values (much higher than the ordinary ones) are reported in Table \ref{Nnu-mir}.
\begin{table}
\begin{tabular}{lrrrrr}
\hline
   \tablehead{1}{r}{b}{T(MeV)}
  & \tablehead{1}{r}{b}{$\mathbf{x=0.1}$}
  & \tablehead{1}{r}{b}{$\mathbf{x=0.3}$}
  & \tablehead{1}{r}{b}{$\mathbf{x=0.5}$}
  & \tablehead{1}{r}{b}{$\mathbf{x=0.7}$}   \\
\hline
5  & 61432 & 761.4 & 101.3 & 28.59\\
0.005 & 74011 & 917.0 & 121.4 & 33.83\\
\hline
\end{tabular}
\caption{Effective $ N_\nu $ in the mirror sector.}
\label{Nnu-mir}
\end{table}

\section{Big Bang Nucleosynthesis}

As we have seen, the presence of the mirror sector can be parametrized in terms of extra DOF number or extra neutrino families; therefore, since the physical processes involved in BBN are not affected by the mirror sector, it is possible to use and modify a standard numerical code to work out the light elements production.

The number of DOF enters the program in terms of neutrino species number; this quantity is a free parameter, but instead of using the same number during the whole BBN process, I use the variable $N_{\nu} (T, x)$ numerically computed following the procedure described in the previous Section.
The only parameter of the mirror sector which affects ordinary BBN is $x$; the baryonic ratio $\beta$ does not induce any changes on the production of ordinary nuclides, but it plays a crucial role for the mirror nuclides production.

In Table \ref{tab-bbn-ord} I report the final abundances (mass fractions) of the light elements $^4He$, $D$, $^3He$ and $^7Li$ produced in the ordinary sector at the end of BBN process (at $T \sim 8\cdot 10^{-4}$ MeV) for several $x$ values and compared with a standard scenario.
We can easily infer that for $x < 0.3$ the light element abundances do not change more than a few percent, and the difference between the standard and $x=0.1$ is of order $10^{-4}$ or less.

\begin{table}
\begin{tabular}{lrrrrrr}
\hline
  & \tablehead{1}{r}{b}{standard}
  & \tablehead{1}{r}{b}{$\mathbf{x=0.1}$}
  & \tablehead{1}{r}{b}{$\mathbf{x=0.3}$}
  & \tablehead{1}{r}{b}{$\mathbf{x=0.5}$}
  & \tablehead{1}{r}{b}{$\mathbf{x=0.7}$}   \\
\hline
$^4He$ & 0.2483 & 0.2483 & 0.2491 & 0.2538 & 0.2675\\
$D/H \: (10^{-5})$ & 2.554 & 2.555 & 2.575 & 2.709 & 3.144\\
$^3He/H \: (10^{-5})$ & 1.038 & 1.038 & 1.041 & 1.058 & 1.113\\
$^7Li/H \: (10^{-10})$ & 4.549 & 4.548 & 4.523 & 4.356 & 3.871\\
\hline
\end{tabular}
\caption{Light elements produced in the ordinary sector.}
\label{tab-bbn-ord}
\end{table}

Even mirror baryons undergo nucleosynthesis via the same physical processes than the ordinary ones, thus we can use the same numerical code also for the M nucleosynthesis.
Mirror BBN is affected also by the second mirror parameter, that is the M baryon density (introduced in terms of the ratio $\beta = \Omega'_b / \Omega_b \sim 1 \div 5$), which raises the baryon to photon ratio $\eta' = \beta x^{-3} \eta$.

The results are reported in Table \ref{tab-bbn-mir}, which is the analogous of Table \ref{tab-bbn-ord} but for a mirror sector with $\beta$ = 5.
We can see that BBN in the mirror sector is much more different from the standard than the BBN in the ordinary sector. This is a consequence of the high ordinary contribution to the number of total mirror DOF, which scales as $\sim x^{-4}$ (while in the ordinary sector the mirror contribution is almost insignificant, since it scales as $\sim x^4$).
Hence, in this case the M helium abundance should be much larger than that of the O helium, and for $x <0.5$ the M helium gives a dominant mass fraction of the dark matter of the Universe.
This is a very interesting feature, because it means that mirror sector can be a helium dominated world, with important consequences on star formation and evolution \cite{Berezhiani:2005vv}, and other related astrophysical aspects.
In particular, I recall the attention of the reader on the fact that the predicted dark matter composition, dominated by mirror helium, is exactly what is required by the proposed interpretation of the DAMA/LIBRA annual modulation signal in terms of mirror dark matter \cite{Bernabei:2008yi,Foot:2008nw}.

\begin{table}
\begin{tabular}{lrrrrrr}
\hline
  & \tablehead{1}{r}{b}{$\mathbf{x=0.1}$}
  & \tablehead{1}{r}{b}{$\mathbf{x=0.3}$}
  & \tablehead{1}{r}{b}{$\mathbf{x=0.5}$}
  & \tablehead{1}{r}{b}{$\mathbf{x=0.7}$}   \\
\hline
$^4He$ & 0.8051 & 0.6351 & 0.5035 & 0.4077\\
$D/H \: (10^{-5})$ & $1.003 \cdot 10^{-7}$ & $4.838 \cdot 10^{-4}$ & $6.587 \cdot 10^{-3}$ & $3.279 \cdot 10^{-2}$\\
$^3He/H \: (10^{-5})$ & $0.3282$ & $0.3740$ & $0.4172$ & $0.4691$\\
$^7Li/H \: (10^{-10})$ & $1.996 \cdot 10^{3}$ & $3.720 \cdot 10^{2}$ & $1.535 \cdot 10^{2}$ & $0.7962 \cdot 10^{2}$\\
\hline
\end{tabular}
\caption{Light elements produced in the mirror sector ($\beta$ = 5).}
\label{tab-bbn-mir}
\end{table}

Considering the different possible non-standard BBN scenarios, it would be of relevance a careful inspection to the effects of the presence of mirror particles on the predicted abundances of light elements.
A more extended investigation in this sense is required in order to evaluate if it could help solving the still present problems on standard BBN due to the lithium anomalies.

\section{Structure formation}

The important moments for the structure formation are related to the matter-radiation equality (MRE) and to the matter-radiation decoupling (MRD) epochs. 
The MRE occurs at the redshift
\begin{equation} \label{z-eq} 
1+z_{\rm eq}= {{\Omega_m} \over {\Omega_r}} \approx 
 2.4\cdot 10^4 {{\Omega_{m}h^2} \over {1+x^4}} ~~.
\end{equation}
Therefore, for $x\ll 1$ it is not altered by the additional relativistic component of the M sector.
The mirror MRD temperature $T'_{\rm dec}$ can be calculated following the same lines as in the O one, obtaining $T'_{\rm dec} \approx T_{\rm dec}$, and hence 
\begin{equation} \label{z'_dec}
1+z'_{\rm dec} \simeq x^{-1} (1+z_{\rm dec}) 
\simeq 1100 \; x^{-1} ~~,
\end{equation}
so that the MRD in the M sector occurs earlier than in the O one. 

Moreover, for values $ x < x_{\rm eq} \simeq 0.046 \, \left( \Omega_{m} h^2 \right)^{-1}$, the mirror photons would decouple yet during the radiation dominated period. 
This critical value plays an important role in our further considerations, where we distinguish between two cases: $x > x_{\rm eq}$ and $x < x_{\rm eq}$. 
For typical values of $\Omega_{m} h^2$ we obtain $x_{\rm eq} \simeq 0.3$.

The relevant scale for gravitational instabilities is the mirror Jeans mass, defined as the minimum scale at which, in the matter dominated epoch, sub-horizon sized perturbations start to grow. 
In the case $x > x_{\rm eq}$ (where the mirror decoupling happens after the matter-radiation equality) its maximum value is reached just before the M decoupling, and is expressed in terms of the O one as
\begin{equation}
M_{\rm J,max}' \approx 
  \beta^{-1/2} \left( { x^4 \over {1 + x^4} } \right)^{\rm 3/2} 
  \cdot M_{\rm J,max} ~~,
\end{equation}
which, for $\beta \geq 1$ and $x < 1$, means that the Jeans mass for the M baryons is lower than for the O ones, with implications for the structure formation.
If, e.g., $ x = 0.6 $ and $ \beta = 2 $, then $ M_{\rm J}' \sim 0.03 \; M_{\rm J} $. 
We can also express the same quantity in terms of $ \Omega_b $, $ x $ and $ \beta $, in the case that all the dark matter is in the form of M baryons, as
\begin{equation} \label{mj_mir_1}
M_{\rm J}'(a_{\rm dec}') \approx 
  3.2 \cdot  10^{14} M_\odot \;
  \beta^{-1/2} ( 1 + \beta )^{-3/2} \left( x^4 \over {1+x^4} \right)^{3/2} 
  ( \Omega_{\rm b} h^2 )^{-2} ~~.
\end{equation}
For the case $ x < x_{\rm eq} $, the mirror decoupling happens before the matter-radiation equality.
In this case we obtain for the highest value of the Jeans mass just before decoupling the expression
\begin{equation}
 \label{mj_mir_2}
M_{\rm J}'(a_{\rm dec}') \approx 
  3.2 \cdot  10^{14} M_\odot \; 
  \beta^{-1/2} ( 1 + \beta )^{-3/2} 
  \left( x \over x_{\rm eq} \right)^{3/2} \left( x^4 \over {1+x^4} \right)^{3/2} 
  ( \Omega_{\rm b} h^2 )^{-2} ~~.
\end{equation}
In case $ x = x_{\rm eq} $, the expressions (\ref{mj_mir_1}) and (\ref{mj_mir_2}), respectively valid for $ x \ge x_{\rm eq} $ and $ x \le x_{\rm eq} $, are coincident, as we expect.
If we consider the differences between the highest mirror Jeans mass for the particular values $ x = x_{\rm eq}/2 $, $ x = x_{\rm eq} $ and $ x = 2 x_{\rm eq} $, we obtain the following relations:
\begin{eqnarray}
M_{\rm J,max}'(x_{\rm eq}/2) &\approx& 0.005 \: M_{\rm J,max}'(x_{\rm eq})
~~; \nonumber\\
M_{\rm J,max}'(2x_{\rm eq}) &\approx& 64 \: M_{\rm J,max}'(x_{\rm eq}) ~~.
\end{eqnarray}
Density perturbations in M baryons on scales $ M \ge M'_{\rm J,max}$, which enter the horizon at $z\sim z_{\rm eq}$, undergo uninterrupted linear growth. 
Perturbations on scales $ M \le M'_{\rm J,max}$ start instead to oscillate after they enter the horizon, thus delaying their growth till the epoch of M photon decoupling.

As occurs for perturbations in the O baryonic sector, also the M baryon density fluctuations should undergo the strong collisional Silk damping around the time of M recombination, so that 
the smallest perturbations that survive the dissipation will have the mass 
\begin{equation} \label{ms_m}
M'_S \sim [f(x) / 2]^3 (\beta \, \Omega_b h^2)^{-5/4} 10^{12}~ M_\odot ~~,
\end{equation}
where $f(x)=x^{5/4}$ for $x \geq x_{\rm eq}$, and $f(x) = (x/x_{\rm eq})^{3/2} x_{\rm eq}^{5/4}$ for $x \leq x_{\rm eq}$. 
For $x\sim x_{\rm eq}$ we interestingly obtain $ M'_S \approx 10^7 (\Omega_b h^2)^{-5} M_\odot \sim 10^{10} M_\odot $, a typical galaxy mass.

\section{Cosmic Microwave Background and Large Scale Structure}

In order to obtain quantitative predictions I computed numerically the evolution of scalar adiabatic perturbations in a flat Universe in which a significant fraction of mirror dark matter is present at the expenses of diminishing the cold dark matter (CDM) contribution and maintaining constant $\Omega_m$. 

\begin{figure}\label{cmblss}
  \includegraphics[height=.5\textheight]{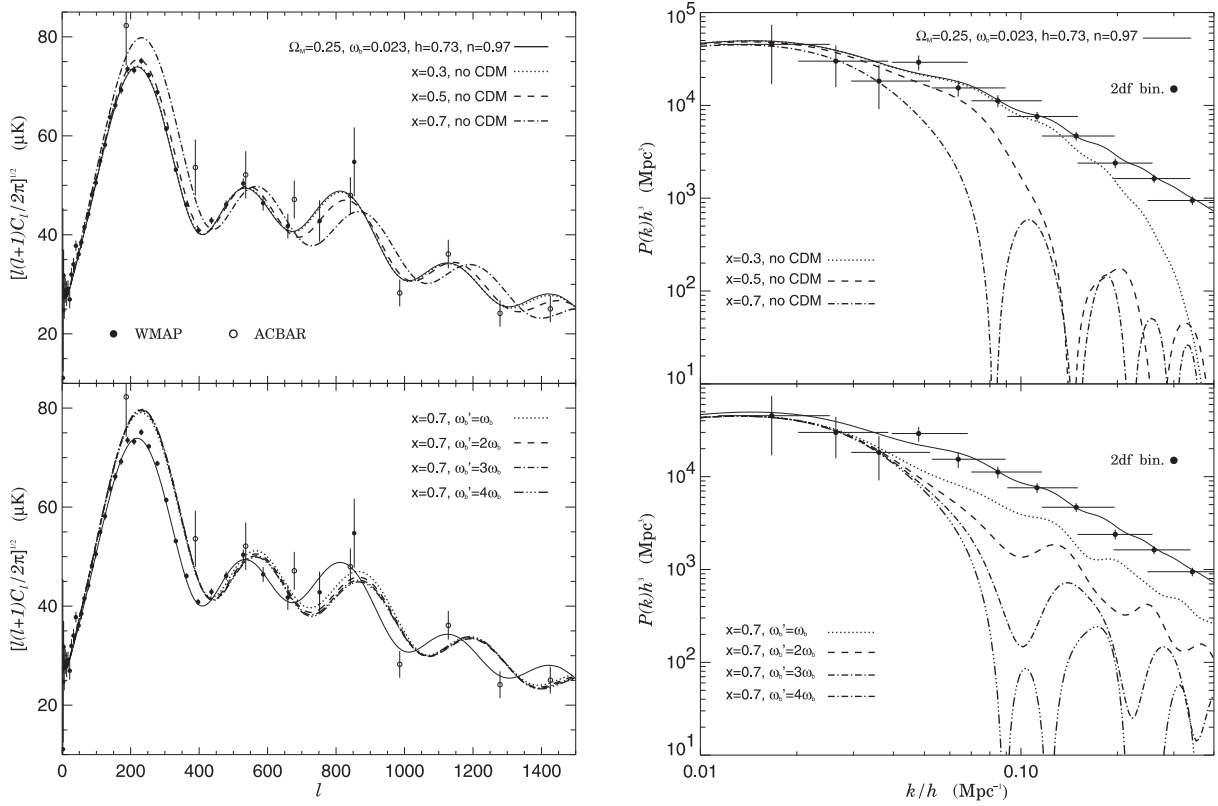}
  \caption{CMB ({\it left}) and LSS ({\it right}) power spectra for different values of $x$ and $\omega_{\rm b}'$, as compared with a reference standard model (solid line) and with observations.
Models where dark matter is entirely due to M baryons (no CDM) are plotted in {\it top panels} for $x = 0.3, 0.5, 0.7$, while models with mixed CDM + M baryons ($\beta=1,2,3,4$ ; $x=0.7$) in {\it bottom panels.} }
\end{figure}

We have chosen a ``reference cosmological model'' with the following set of parameters:
$ \omega_{b} = \Omega_{b} h^2 = 0.023, ~ 
\Omega_{ m} = 0.25, ~ 
\Omega_{\Lambda} = 0.75 , ~ 
n_{\rm s} = 0.97, ~ h = 0.73 $. 
The dependence of the CMB and LSS power spectra on the parameters $x$ and $\beta$ is shown in Fig. \ref{cmblss}.
The predicted CMB spectrum is quite strongly dependent on the value of $x$, and it becomes practically indistinguishable from the CDM case for $x < x_{\rm eq} \approx 0.3$.
However, the effects on the CMB spectrum rather weakly depend on the fraction of mirror baryons. 
As a result of the oscillations in M baryons perturbation evolution, one observes oscillations in the  LSS power spectrum; their position clearly depends on $x$, while their depth depends on the mirror baryonic density.
Superimposed to oscillations one can see the cut-off in the power spectrum due to the aforementioned mirror Silk damping.

In the same Figure our predictions can be compared with the observational data in order to obtain some general bound on the mirror parameters space.

\begin{itemize}
\item The present LSS data are compatible with a scenario where all the dark matter is made of mirror baryons only if we consider enough small values of $ x $: $ x \lsim 0.3 \approx x_{\rm eq} $.
\item High values of $ x $, $ x > 0.6 $, can be excluded even for a relatively small amount of mirror baryons. 
In fact, we observe relevant effects on LSS and CMB power spectra down to values of M baryon density of the order $ \Omega'_b \sim \Omega_b $. 
\item Intermediate values of $ x $, $ 0.3 < x < 0.6 $, can be allowed if the M baryons are a subdominant component of dark matter, $ \Omega_b \lsim \Omega_b' \lsim \Omega_{CDM} $. 
\item For small values of $ x $, $ x < 0.3 $, 
the M baryons and the CDM scenarios are indistinguishable as far as the CMB and the linear LSS power spectra are concerned.
In this case, in fact, the mirror Jeans and Silk lengths, which mark region of the spectrum where the effects of mirror baryons are visible, decrease to very low values, which undergo non linear growth from relatively large redshift. 
\end{itemize}

Thus, with the current experimental accuracy, we can exclude only models with high $ x $ and high $ \Omega_b' $.

\section{Photon--mirror-photon kinetic mixing and DAMA experiment}

We have so far considered only gravitational interactions between O and M particles.
Nevertheless, I mentioned in the Introduction the existence of other kind of weak interactions linking the two sectors.
The most important of these is realized if O(M) particles couple to the M(O) particles via renormalizable photon--mirror-photon kinetic mixing \cite{Foot:1991kb}.
The mixing term in the Lagrangian is 
\begin{eqnarray}
{\cal L}_{mix} = {\epsilon \over 2}F^{\mu \nu} F'_{\mu \nu} ~~,
\end{eqnarray}
where $F^{\mu \nu} = \partial^{\mu} A^{\nu} - \partial^{\nu}A^{\mu}$ and $F'^{\mu \nu} = \partial^{\mu} A'^{\nu} - \partial^{\nu}A'^{\mu}$.
This mixing enables O(M) charged particles to couple to M(O) photons with charge $\pm \epsilon e$.
In the last years it has acquired more and more importance, since it has been shown that mirror dark matter is able to explain the positive dark matter signal obtained in the DAMA/Libra experiment \cite{Bernabei:2008yi}, while also being consistent with the null results of the other direct detection experiments \cite{Foot:2008nw}.
This interpretation in terms of mirror dark matter requires $\epsilon \sim 10^{-9}$, which is consistent with current laboratory and astrophysical constraints.
The implications of such mixing for the early Universe have been recently studied in ref. \cite{Ciarcelluti:2008qk}, where it has been shown in particular its consistency with constraints from ordinary BBN as well as more stringent constraints from CMB and LSS considerations.

Assuming an effective initial condition $T' \ll T$, we can estimate the evolution of $T'/T$ in the early Universe, and also estimate the $He'$ abundance.
Photon--mirror-photon kinetic mixing can populate the mirror sector in the early Universe, via the process $e^+ e^- \to {e^+}' {e^-}'$.
The ${e^+}',{e^-}'$ will interact with each other via mirror weak and mirror electromagnetic interactions, populating the $\gamma', \nu_e', \nu_\mu', \nu_\tau'$, and thermalizing to a common
mirror sector temperature $T'$.

Employing the usual expressions for the densities of relativistic fermions in thermal equilibrium and integrating with the initial condition $T' = 0$ at $T = T_i$, we derive the evolution of the $T'/T$ ratio as a decreasing function of $T$
\begin{eqnarray}
{T' \over T} 
\simeq 0.52 \cdot 10^4 \,\epsilon^{1/2} \left({MeV \over T} - {MeV \over T_i} \right)^{1/4}
\simeq 0.164 ~\epsilon_{-9}^{1/2} \left({1 \over T(MeV)} \right)^{1/4} ~~.
\label{T'overT}
\end{eqnarray}
The last equivalence is obtained considering $T_i \gg 1 MeV$ and defining $\epsilon_{-9} = \epsilon /10^{-9}$.
Clearly in all these computations we have neglected the change of the $T'/T$ ratio due to the $e^+$-$e^-$ annihilation processes in both sectors (for a detailed study of this effect see ref. \cite{Ciarcelluti:2008vs}), and the transfer from the mirror sector to the ordinary one (since $T'<T$).

Note that the $T'/T$ ratio freezes out when $T \lsim 2\,m_e$   since the number density of $e^{\pm}$ becomes Boltzmann suppressed and the process $e^+ e^- \to {e^+}' {e^-}'$ can no longer effectively heat the mirror sector.
In addition, after ${e^+}'$-${e^-}'$ annihilation the effective number of mirror degrees of freedom decreases, and thus we estimate:
\begin{eqnarray}
\frac{T'}{T} \simeq 0.2~\epsilon^{1/2}_{-9}\
~~~~{\rm for} ~~\ T \lsim 1 {\rm MeV} ~~.
\label{new4}
\end{eqnarray}
Constraints from ordinary BBN suggest that $\delta N_{\nu} \lsim 0.5 \Rightarrow T'/T < 0.6$.
A more stringent constraint arises from CMB and LSS, which suggest \cite{Berezhiani:2003wj,Ciarcelluti:2003wm,Ciarcelluti:2004ip} $T'/T \lsim 0.3$ and implies, from Eq.~(\ref{new4}), that $\epsilon \lsim 3\times 10^{-9}$.
This upper limit is compatible with the value required in order to explain the DAMA annual modulation signal \cite{Foot:2008nw}.

Given the above expression for the $T'/T$ evolution, we can estimate the mirror helium mass fraction as a function of $\epsilon$.
It has already been discussed in ref.\cite{Berezhiani:2000gw,Ciarcelluti:2008vm} that, compared with the ordinary matter sector, we expect a larger mirror helium mass fraction if $T' < T$.
Essentially, this is because the expansion rate of the Universe is faster at earlier times, which implies that the freeze out temperature of mirror weak interactions will be higher than that in the ordinary sector.

The relevant time scales for BBN are the freeze out temperature of weak interactions $T_W$, and the ``deuterium bottleneck'' temperature $T_N$ and time $t_N$.

In the mirror sector, we have the same relations describing the BBN process, except that we change
$T_W \rightarrow T_W'$ and $t_N \rightarrow t'_N$.
Imposing the equality of weak interaction and Hubble rates $\Gamma_W(T_W') =H(T_W')$, we obtain
\begin{eqnarray}
T_W' \simeq B^{-2/7} \,\epsilon^{-4/7} \,T_W^{9/7} ~~,
\label{TW'}
\end{eqnarray}
with  $B=0.74 \cdot 10^{15}$ MeV and where we used $T_W \simeq \left(1.66 ~g^{1/2} \over{G_F^2 ~M_{Pl}}\right)^{1/3}$, as computed from the equation $\Gamma_W(T_W) = H(T_W)$.
The quantity $t_N'$ can be estimated assuming $T_N' \sim T_N$ (since $T_N$ depends only logarithmically on the baryon to photon ratio, that is larger in the M sector):
\begin{eqnarray}
t_N'
\simeq 0.3 ~g^{-1/2}\left[1+\left({T' \over T}\right)^{-4}
\right]^{-1/2}{M_{Pl} \over T_N^2}
\simeq B^{1/2} \epsilon \,T_N^{-1/2}\,t_N ~~,
\label{tN'}
\end{eqnarray}
where we used $t_N \simeq 0.3 ~g^{-1/2}{M_{Pl} \over T_N^2}$ and $B^{-1}\epsilon^{-2}T_N \gg 1$.

Finally, using Eqs.~(\ref{TW'}) and (\ref{tN'}), we can estimate the mirror helium mass fraction:
\begin{eqnarray}
Y_4' \simeq 2\,X_n' (T_W', t_N')
\simeq {2\,exp(-t_N'/\tau) \over 1+exp(\Delta m/T_W')}
\simeq {2\,exp\left[{-B^{1/2} ~T_N^{-1/2} \over \tau} t_N ~\epsilon \right]
\over 1+exp\left[{\Delta m~B^{2/7}\over T_W^{9/7}}~\epsilon^{4/7}\right]} ~~.
\label{YHe'}
\end{eqnarray}
Using the typical values $T_W \simeq 0.8$ MeV, $T_N \simeq 0.07$ MeV, $\Delta m \simeq 1.29$ MeV, $B \simeq 0.74 \cdot 10^{15}$ MeV, $\tau \simeq 886.7$ s, $t_N \simeq 200$ s, we obtain:
\begin{eqnarray}
Y_4' \simeq {2\,exp\left[-2.3 \cdot 10^7 ~\epsilon \right] \over
1+exp\left[0.3 \cdot 10^5 ~\epsilon^{4/7} \right]}
\simeq {2\,exp\left[-2.3 \cdot 10^{-2} ~\epsilon_{-9} \right] \over
1+exp\left[0.22 ~\epsilon_{-9}^{4/7} \right]} ~~.
\end{eqnarray}
Thus, for $\epsilon \simeq 10^{-9}$, we find $Y_4' \simeq 0.87$, that means a mirror sector largely dominated by helium.

\section{Summary}

\begin{figure}\label{resume}
  \includegraphics[height=.56\textheight]{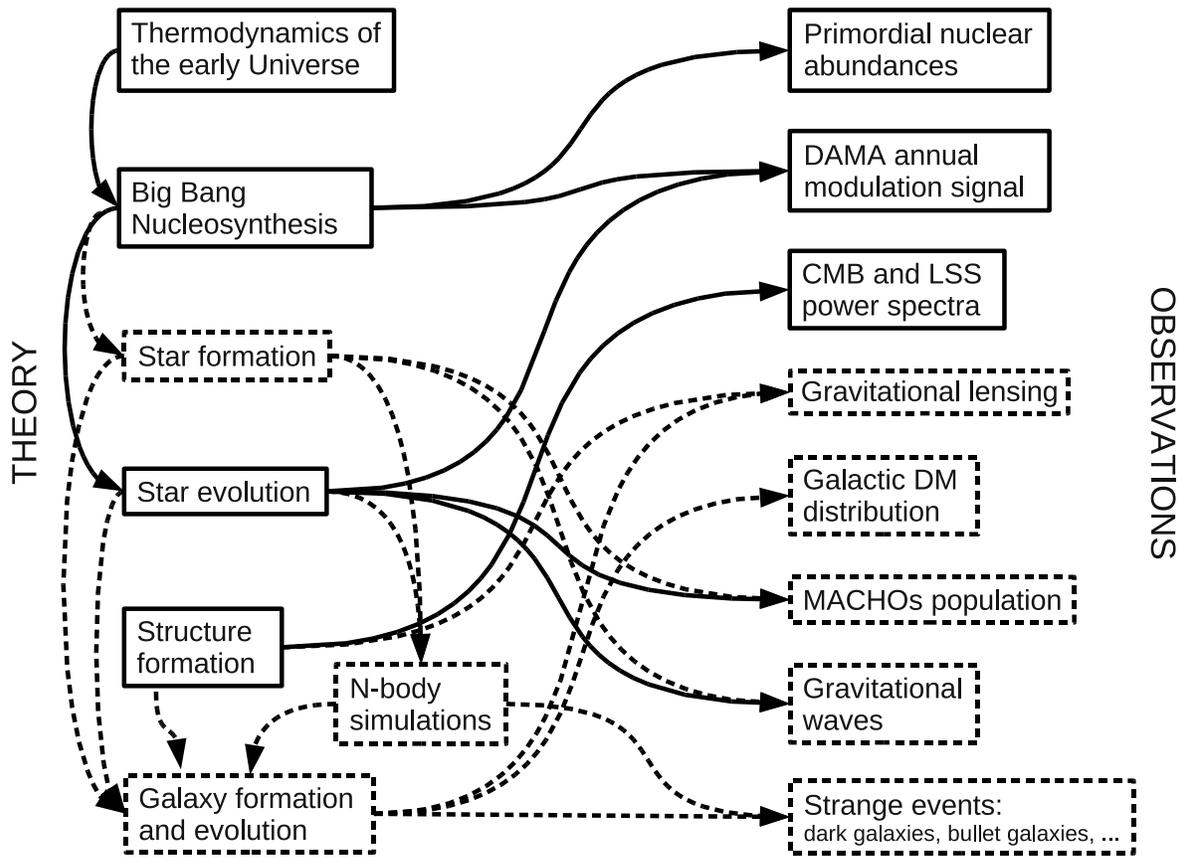}
  \caption{Current status of the astrophysical research with mirror dark matter: solid lines mark what is already done, while dashed ones mark what is still to do. For a more extensive explanation see the text.}
\end{figure}

The current situation of the astrophysical research in presence of mirror dark matter is shown in Figure \ref{resume}, with emphasis on the connections between the different theoretical studies and the predicted observable signatures.
Solid lines mark what is already done (and is compatible with current observational and experimental constraints), while dashed ones mark what is still to do.
For the last point a special importance is covered by star formation and N-body simulation studies, towards which people interested in mirror dark matter should address their efforts in the near future.

At present, we have already investigated the early Universe (thermodynamics and Big Bang nucleosynthesis), and the process of structure formation in linear regime, that permit to obtain predictions, respectively, on the primordial elements abundances, and on the observed cosmic microwave background and large scale structure power spectra.
In addition, we have studied the evolution of mirror dark stars, which, together with the mirror star formation, are necessary ingredients for the study and the numerical simulations of non linear structure formation, and of the formation and evolution of galaxies. 
Furthermore, in future studies they can provide predictions on the observed abundances of MACHOs and on the gravitational waves background.
Using the results of mirror BBN and stellar evolution, we have verified that the interpretation, in terms of photon--mirror-photon kinetic mixing, of the DAMA annual modulation signal is compatible with the inferred consequences on the physics of the early Universe.
Ultimately, we will be able to obtain theoretical estimates of gravitational lensing, galactic dark matter distribution, and strange astrophysical events still unexplained (as for example dark galaxies, bullet galaxy, ...), that can be compared with observations.

Concluding, the astrophysical and experimental tests so far used show that mirror matter is not just a viable, but a promising candidate for dark matter. 
However, we still need to obtain a complete picture of the Mirror Universe.

\begin{theacknowledgments}
This work was supported by the Belgian Science Policy Office Inter University Attraction Pole VI/11 ``Fundamental Interactions''.
\end{theacknowledgments}



\bibliographystyle{aipproc}   

\bibliography{InvUniv2009ciarcelluti}

\IfFileExists{\jobname.bbl}{}
 {\typeout{}
  \typeout{******************************************}
  \typeout{** Please run "bibtex \jobname" to optain}
  \typeout{** the bibliography and then re-run LaTeX}
  \typeout{** twice to fix the references!}
  \typeout{******************************************}
  \typeout{}
 }

\end{document}